\documentclass[onecolumn,prd,superscriptaddress,nofootinbib,notitlepage]{revtex4-1}
\pdfoutput=1
\usepackage[T1]{fontenc}
\usepackage{bm}
\usepackage{subfigure}
\usepackage{amssymb}
\usepackage{graphicx}
\usepackage{amsmath}
\usepackage{tensor}
\usepackage{epstopdf}
\usepackage{extarrows}
\usepackage{xcolor}
\usepackage[colorlinks=true,
              linkcolor=blue,
              urlcolor=blue,
              citecolor=blue
              ]{hyperref}

\usepackage{latexsym}
\usepackage{amsmath,amsfonts,graphicx,accents}
\usepackage{amsbsy}
\usepackage{hyperref}
\usepackage{mathrsfs}
\usepackage{color}
\usepackage{psfrag}
\usepackage{enumerate}
\usepackage{amsmath,amssymb,calc,amsfonts}
\usepackage{latexsym}
\usepackage[utf8]{inputenc}
\usepackage[percent]{overpic}
\usepackage[autostyle=false, style=english]{csquotes}
\MakeOuterQuote{"}
\DeclareFontFamily{U}{rsfs}{}         
\DeclareFontShape{U}{rsfs}{m}{n}{<5> rsfs5 <6><7> rsfs7          %
  <8><9><10><10.95><12><14.4><17.28><20.74><24.88> rsfs10}{}     %
\DeclareMathAlphabet{\mathfs}{U}{rsfs}{m}{n}                     %
\definecolor{indiagreen}{rgb}{0.07, 0.53, 0.03}
\def\beq{\begin{eqnarray}}
\def\eeq{\end{eqnarray}}

\def\={\stackrel{\Delta}{=}}

\begin{document}
\title{Phase-space Path Integral Approach to the Kinetics of Black Hole Phase Transition in Massive Gravity}
\author{C. Fairoos}\email{fairoos.phy@gmail.com}
\affiliation{T. K. M. College of Arts and Science Kollam- 691005, India}
\author{T. K. Safir}\email{stkphy@gmail.com}
\affiliation{T. K. M. College of Arts and Science Kollam- 691005, India}
\author{Deepak Mishra}\email{deepak.mishra@iist.ac.in}
\affiliation{Indian Institute of Space Science and Technology Trivandrum}
 
\begin{abstract}
The dynamics of the state-switching process of black holes in dRGT massive gravity theory is presented using free energy landscape and stochastic Langevin equations. The free energy landscape is constructed using the Gibbons-Hawking path integral method. The black hole phases are characterized by taking its horizon radius as the order parameter. The free energy landscape provides three black hole phases: small, intermediate, and large. The small and large black holes are thermodynamically stable whereas the intermediate one is unstable. The Martin–Siggia–Rose–Janssen–de Dominicis (MSRJD) functional describes the stochastic dynamics of black hole phase transition. The Hamiltonian flow lines are obtained from the MSRJD functional and are used to analyze the stability and the phase transition properties. The dominant kinetic path between different phases is discussed for various configurations of the free energy landscape. We discuss the effect of black hole charge and the graviton mass on the critical behavior of black hole phase transition.
\end{abstract}

 \maketitle
\section{Introduction}\label{intro}
Tools of thermodynamics and statistical mechanics are now widely used to understand black holes. The motivation behind such studies is to establish a connection between gravity, quantum theory, and statistical physics. The discovery of black hole phase transition between Schwarzschild Anti-de-Sitter (AdS) black hole and thermal AdS space (Hawking-Page phase transition) is a remarkable achievement in this pursuit \cite{Hawking:1982dh}. The thermodynamic stability of the black hole demands that the asymptotic structure of space-time should be AdS. Accordingly, black hole phase transitions were extensively investigated for various black hole solutions in AdS space \cite{Chamblin:1999tk,Chamblin:1999hg,Caldarelli:1999xj,Kubiznak:2012wp,Altamirano:2013ane,Cheng:2016bpx}. Such studies were also carried out for black holes in different theories of gravity. Einstein's general theory of relativity (GR) is an effective field theory that describes how gravity works classically. One can modify the corresponding Lagrangian selectively and still obtain a description of gravity that is mathematically consistent and reproduce correct results in appropriate limits.
Interestingly, AdS black holes in such modified gravity theories also exhibit phase transition similar to the van-der-Waal systems \cite{Cai:2013qga, Zou:2014mha,Xu:2014tja,Frassino:2014pha, Hennigar:2016xwd,Hennigar:2017umz,Chabab:2018zix}. In this paper, we will be considering black hole phase transition in a modified description of gravity called massive gravity theory. The basic idea of massive gravity theory, introduced in 1939, was to put a non-zero mass to the spin-2 gravitons. Many modifications were made since then to obtain the most consistent version proposed by de Rham, Gabadadze, and Tolley (dRGT) in 2010 \cite{deRham:2010kj}. A complete history of the massive gravity theory can be found in \cite{deRham:2014zqa}. The dRGT massive gravity theory modifies GR in large scales (IR limit) and is ghost-free \cite{Hassan:2011hr}.  This theory can explain the accelerated expansion of the universe without evoking dark energy \cite{Akrami:2012vf,Hassan:2011tf}. Also, dRGT massive gravity theory is consistent with recent observational findings of LIGO collaboration\cite{LIGOScientific:2017bnn}. \\

The free energy landscape is a powerful tool to study phase transition in many physical systems \citep{Frauenfelder:1991fs, Goldenfeld:1992qy, Wang2015LandscapeAF}. One can explore the construction of the free energy landscape for black hole systems where a black hole is considered as the macroscopic emergent state of microscopic degrees of freedom. A state of the black hole is described by its horizon radius, which serves the role of order parameter of the theory. The free energy landscape is used to study the thermodynamic stability and the phase transition of black hole states in given gravity theory. Here, the free energy is obtained from the basic thermodynamic relationship between internal energy, entropy, and temperature. In \cite{Li:2020nsy, Wei:2020rcd, Safir:2023thg} the Gibbs free energy is used to construct the free energy landscape to study phase transition characteristics of black holes in GR and other modified gravity theories. The generalized free energy can also be treated as a potential defined on the thermodynamic parameter space and study Duan's $\phi$ field topological current theory \cite{Duan:1979ucg, Duan:1984ws}. In the thermodynamic topological analysis, black hole solutions are considered as topological thermodynamic defects. One of the aspects of topological studies is to use the topological methods to study the phase structure, critical phenomena, and stability of the black hole system. For example, the value of the winding number constructed from the potential reflects the local stability. A positive (negative) winding number corresponds to a locally thermodynamically stable (unstable) black hole solution. Another quantity in these analyses is the topological number according to which all black hole solutions are classified into three categories \cite{Wei:2022dzw}. A detailed discussion on topological analysis of black holes in massive gravity theory is given in \cite{Fairoos:2023jvw}. Interestingly, one can also study the black hole phase transition using the tools of topological theory. In this set-up,  a first-order phase transition between two different black hole phases is interpreted as the interchange of winding numbers between the defects as a result of some action at a distance \cite{Fan:2022bsq,Fairoos:2023hkk}.\\

The deterministic evolution of black hole states can be qualitatively explained using the free energy landscape. However, as the phase switching phenomena occur due to the stochastic fluctuations in the thermal bath, one should exploit the tools of non-equilibrium statistical mechanics to address the dynamics of phase transition. The dynamics of state-switching processes are governed by the stochastic Langevin equation or equivalently Fokker-Planck equation for probabilistic evolution. In \cite{Li:2024hje}, the free energy landscape for black holes in GR and Einstein-Gauss-Bonnet gravity theory was constructed from the gravitational path integral formalism, and the dynamics of black hole phase switching processes were studied using MSRJD functional. Motivated by this, we will construct the Gibbs free energy for black holes in dRGT massive gravity from the corresponding Euclidean gravitational action and study the dynamics of phase transition using Hamiltonian flow lines obtained from MSRJD functional. \\

In the coming section, we will briefly outline the basic thermodynamic properties of black holes in dRGT massive gravity theory followed by an explicit calculation of the Gibbs free energy using the path integral approach. In \ref{second}, we will derive the MSRJD functional from the over-damped Langevin equation. We will obtain the Hamiltonian equations of motion from the MSRJD functional and the dynamics of the phase switching process are studied in \ref{third}. A summary of the work is presented in \ref{dis}.

\section{Generalized Free Energy of \lowercase{d}RGT Massive Gravity}\label{thermodynamics}

We briefly outline the thermodynamic structure of black hole solutions in dRGT non-linear massive gravity theory in AdS space, described by the action \cite{deRham:2010ik,deRham:2010kj,Vegh:2013sk,Cai:2014znn},
\begin{equation}
S=\int d^4 x \sqrt{-g} \left[ \frac{1}{16 \pi}\left[ R+\frac{6}{L^2}+m^2\sum_{i=1}^4 c_i \ \mathcal{U}_i(g,f)\right ]-\frac{1}{4 } F_{\mu \nu}F^{\mu \nu}\right],
\end{equation} 
 where $F_{\mu \nu}=\partial_\mu A_\nu-\partial_\nu A_\mu $ is the electromagnetic field tensor with vector potential $A_\mu$, $L$ is the AdS radius, $m$ is related to the graviton mass, and $c_i$ are coupling parameters. The symmetric tensor $f_{\mu \nu}$ is the auxiliary reference metric coupled to the space-time metric $g_{\mu \nu}$ and is needed to assign a nonzero graviton mass. Each choice of the reference metric gives different massive gravity theories \cite{Hassan:2011tf}. Graviton interaction terms are represented by symmetric polynomials $\mathcal{U}_i$ \cite{Cai:2014znn}, where,
 \begin{eqnarray*}
\mathcal{U}_1 &= &\frac{2}{r},\\
\mathcal{U}_2 & = &\frac{2}{r^2},\\
\mathcal{U}_3  = \mathcal{U}_4 &=& 0.
\end{eqnarray*}
The above conditions imply that one can put the coupling parameters $c_3=c_3=0$. Note that this condition is true only in four spacetime dimensions. In higher dimensions, the dRGT massive gravity theory exhibits a rich thermodynamic structure as there are more coupling parameters. The spherically symmetric space-time metric describing the above action is of the form,

\begin{equation}\label{metric}
ds^2=-f(r)dt^2+\frac{1}{f(r)}dr^2+r^2 d\Omega^2,
\end{equation}
where the metric function $f(r)$ is given by \cite{Cai:2014znn},

\begin{eqnarray}\label{metric_fn}
f(r) &=& 1 - \frac{ M}{ r}+\frac{r^2}{L^2}  + \frac{Q^2}{4 r^{2}}
+ m^2\left(\frac{c_1}{2} r 
+  c_2 \right).
\end{eqnarray}
Here, $M$ and $Q$ are related to black hole mass and charge, respectively. The event horizon ($r_h$) is determined by the largest root of the equation $f(r)|_{r=r_h}=0$. The mass, temperature, and entropy of the black hole in Einstein-dRGT gravity coupled to a non-linear electromagnetic field can be expressed in terms of the horizon radius  $r_h$ and the cosmological constant as following:
\begin{eqnarray} \nonumber
M&=&\frac{r_h}{2}\left(1+\frac{r_h^2}{L^2}  + \frac{Q^2}{4 r_h^2}+m^2 \left(\frac{c_1}{2} r_h + c_2\right)\right),\\ \nonumber
T_H&=& \left( \frac{3r_h}{4 \pi L^2}+\frac{1+c_2m^2}{4\pi r_h	}-\frac{Q^2}{16\pi r_h^3} +\frac{c_1 m^2}{4\pi}\right),\\ \label{3basics}
S&=& \pi r_h^2.
\end{eqnarray} 

As explained before, the black hole phase transition is studied using the free energy which can be obtained from the gravitational partition function. According to Gibbons-Hawking, the partition function of a canonical ensemble can be expressed in the form of path integral, given by \cite{Gibbons:1976ue,Hawking:1976ja}, 

\begin{equation}\label{path1}
Z_{gravity}(\beta) = \int D[g] \ e^{-I_E[g]}.
\end{equation}
Here, $I_E[g]$ is the Euclidean gravitational action, and $\beta$ is the period of Euclidean time coordinate. The Euclidean time $\tau=i t$ is introduced by a Wick rotation in the complex $t$ plane. The corresponding Euclidean metric looks like,
\begin{equation}\label{Emetric}
ds^2=f(r)d\tau^2+\frac{1}{f(r)}dr^2+r^2 d\Omega^2.
\end{equation}
In Eq. \ref{path1}, the measure of functional integral $ D[g]$ should be taken on all the Euclidean gravitational configurations that satisfy a given set of boundary conditions. However, one can exploit the saddle point approximation and evaluate the the functional integral over the fluctuating black hole only to obtain \cite{Gibbons:1976ue},
\begin{equation}
Z_{gravity}(\beta) \ \simeq e^{-I_E[g]}.
\end{equation}
 For a well-defined theory of gravity, the Euclidean action contains the following contributions:\begin{equation}\label{contri}
I_E = I_{Bulk} + I_{Surf}+I_{Count}.
\end{equation}
 Now, the bulk action for Einstein-dRGT gravity coupled to a non-linear electromagnetic field in  AdS space comes from the Einstein-Hilbert Lagrangian with the interaction terms added, i.e.,
 \begin{equation}\label{action}
I_{Bulk}=  - \frac{1}{16 \pi} \int d^{4} x \sqrt{g} \left[ {R}+\frac{6}{L^2}+\frac{2m^2}{r}\left(c_1+\frac{c_2}{r}\right)-\frac{1}{4 } F_{\mu \nu}F^{\mu \nu}\right].
\end{equation} 

The surface term has both geometric and electromagnetic terms given by,
\begin{equation}
I_{Surf} = -\frac{1}{8\pi}\int_{\partial {\mathcal{M}}} d^3x \sqrt{h}\left(K + 2 n_\mu F^{\mu \nu} A_\nu\right).
\end{equation}
The integration is over the hypersurface $\partial {\mathcal{M}}$ at a large radius. $K$ is the trace of extrinsic curvature of the hypersurface having induced metric $h_{\mu \nu}$ and $n^\mu$ is the outward pointing unit normal vector to the hypersurface, i.e., $h_{\mu \nu} = g_{\mu \nu} - n_\mu n_\nu$. The counter term for dRGT massive gravity theory, sufficient to cancel the divergences in the action and the boundary stress energy tensor, is given by \cite{Cao:2015cza},
\begin{equation}
 {I}_{Count} = \frac{1}{8 \pi}\int_{\partial \mathcal{M}} d^3x \sqrt{h}\Big[\frac{4}{L^2}+ 2 \mathcal{R} + m^2\left(2\frac{c_1}{r_c}+4\frac{c_2}{r_c^2}\right)\Big]^{\frac{1}{2}}.
 \end{equation}
Here, $\mathcal{R} $ is the Riemann curvature scalar on $\partial {\mathcal{M}}$. The divergences in the gravity action come from the integration over the near-boundary region of the AdS spacetime. In the spirit of AdS/CFT, adding the counter terms is equivalent to the renormalization schemes in the field theory. A prescription to obtain the local counter-terms for gravitational action in both asymptotically flat and AdS spacetimes are given in \cite{Kraus:1999di}. In the following, we will explicitly calculate each term in Eq. \ref{contri} for dRGT massive gravity theory characterized by the metric in Eq. \ref{Emetric}.


\subsection{The Bulk Term}
Note that in the original prescription by Hawking and Gibbons, the period $\beta$ was fixed and set to equal to the inverse temperature of the thermal bath. However, in this paper, $\beta$ is taken arbitrary. Therefore, the Euclidean metric given in Eq. \ref{Emetric} suffers a conical singularity $\Sigma$ at the horizon. The bulk action contribution from the conical singularity is given by \cite{Fursaev:1995ef},
\begin{equation}
-\frac{1}{16 \pi} \int_{\Sigma} d^4 x \sqrt{g} {R} = - \left(1-\frac{\beta}{\beta_H}\right) \pi  r_h^2,
\end{equation}
where $\beta_H = 1/T_H$ is the inverse Hawking temperature as given in Eq. \ref{3basics}. Now, the total contribution of bulk action becomes,
\begin{eqnarray}
\mathcal{I}_{bulk}
&=&- \left(1-\frac{\beta}{\beta_H}\right) \pi  r_h^2- \frac{1}{16 \pi} \int_{\mathcal{M}/\Sigma} d^4 x \sqrt{g} \left[ {R}+\frac{6}{L^2}+\frac{2m^2}{r}\left(c_1+\frac{c_2}{r}\right) -\frac{1}{4 } F_{\mu \nu}F^{\mu \nu} \right]\\
&=& - \left(1-\frac{\beta}{\beta_H}\right) \pi  r_h^2 + \frac{1}{16 \pi} \int_{\mathcal{M}/\Sigma} d^4 x \sqrt{g} \left( \frac{6}{L^2}+ \frac{Q^2}{2 r^4} + \frac{c_1 m^2}{r}\right).
\end{eqnarray}
Here, ${\mathcal{M}/\Sigma}$ represents the part of spacetime with conical singularity cut-off. The above integration can be performed explicitly to obtain,
\begin{eqnarray} \nonumber
\frac{1}{16 \pi} \int_{\mathcal{M}/\Sigma} d^4 x \sqrt{g} \left( \frac{6}{L^2}+ \frac{Q^2}{2 r^4} + \frac{c_1 m^2}{r}\right)
&=& \frac{\beta}{4}\int_{r_h}^{r_c} r^2 dr \left( \frac{6}{L^2}+ \frac{ Q^2}{2 r^4} + \frac{c_1 m^2}{r}\right),
\\ 
&=&\frac{\beta}{2} \Big[ \frac{1}{L^2}\left(r_c^3-r_h^3\right) +\frac{c_1 m^2}{4}\left(r_c^2-r_h^2\right) - \frac{Q^2}{4}\left(\frac{1}{r_c}-\frac{1}{r_h}\right)\Big],
\end{eqnarray}
where $r_c$ denotes the radius of the boundary $\partial \mathcal{M}$. Note that $r_c$ is arbitrary and will be set to infinity in the end.
\subsection{ Gibbons-Hawking-York Surface Term}
The Gibbons-Hawking-York surface term is added to the action for non-null boundaries so that the variational principle becomes well-posed. The outward pointing unit vector on $\partial \mathcal{M}$ with a fixed radius $r_c$ is defined as,
\begin{equation*}
n^\mu = \left(0, \sqrt{f(r)}, 0, 0\right).
\end{equation*}
The trace of extrinsic curvature $K$ is given by,
\begin{equation}
K = \nabla_\mu n^\mu = \frac{f'}{2\sqrt{f}}+\frac{2}{r}\sqrt{f},
\end{equation}
where $f'=\frac{\partial f}{\partial r}$. Now the required surface term is readily obtained as,
\begin{eqnarray}
-\frac{1}{8\pi}\int_{\partial \mathcal{M}} d^3x\sqrt{h} K 
 = -\frac{\beta}{2} \Big[ \frac{3 r_c^3}{L^2}-3M + \frac{Q^2}{r_c} + 2(1+c_2 m^2)r_c+\frac{5}{4} m^2 c_1 r_c^2\Big].
 \end{eqnarray}
 \subsection{Electromagnetic Surface Term}
 The surface term for the electromagnetic field is,
 \begin{equation}
 -\frac{1}{4\pi}\int_{\partial \mathcal{M}} d^3x \sqrt{h} n_{\mu} F^{\mu \nu}A_\nu = \frac{\beta Q^2}{r_c},
 \end{equation}
 where we have used $A^\mu = (-\frac{Q}{r}, 0, 0, 0)$. The contribution from the electromagnetic surface term is the same for a charged black hole both in GR as well as in massive gravity as expected.
 \subsection{Counter Term}
\begin{figure}[hbt!]
    \centering
    \includegraphics[width=0.4\textwidth]{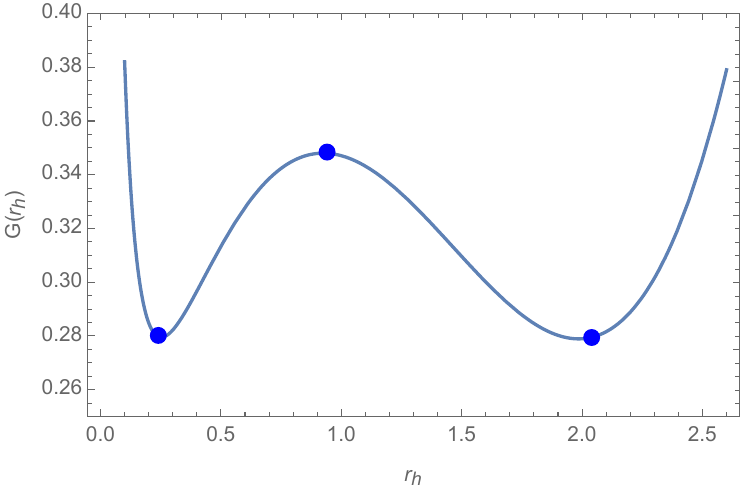}\;\;\;
    \includegraphics[width=0.4\textwidth]{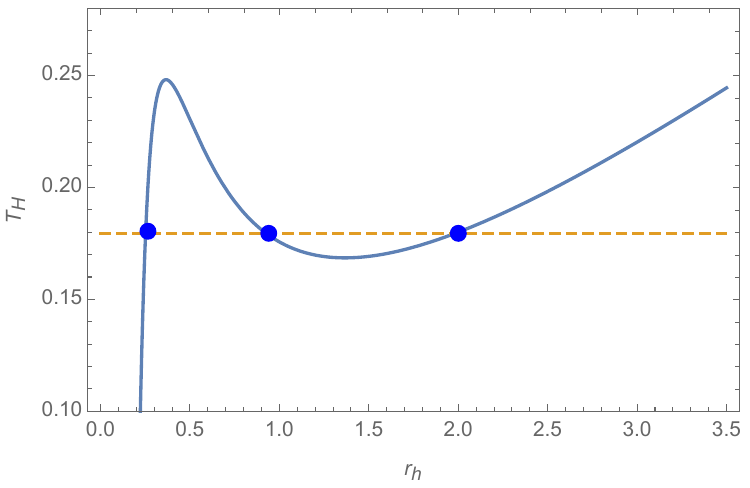}
      \caption{ \textit{On the left:} At the phase transition temperature the free energy takes the double-well shape and the free energy of small and large black holes are equal.   \textit{On the right:} The behaviour of Hawking temperature against the order parameter. The dashed horizontal line represents the phase transition temperature. Numerical values used: $Q=0.5, P =0.03, m=1,c_1=0.02$, and $c_2=0.5$}
  \label{fig_1}
\end{figure}

 The solution independent counter term added to the action so that the action becomes finite is given by \cite{Cao:2015cza},
 \begin{eqnarray}
 \mathcal{I}_{Count} &=& \frac{1}{8 \pi}\int_{\partial \mathcal{M}} d^3x \sqrt{h}\Big[\frac{4}{L^2}+ 2 \mathcal{R} + m^2\left(2\frac{c_1}{r_c}+4\frac{c_2}{r_c^2}\right)\Big]^{\frac{1}{2}}\\
 &=& \frac{\beta}{2}\Big[ \frac{2 r_c^3}{L^2}+c_1 m^2 r_c^2 - 2M +2\left(1+c_2m^2\right)r_c + \frac{Q^2}{r_c} + \mathcal O\left(\frac{1}{r_c}\right)^3\Big].
 \end{eqnarray}
We neglect the higher order terms in $1/r_c$ because we will send $r_c$ to infinity. In this limit, the total Euclidean action can be calculated by summing over all contributions detailed above,

 \begin{eqnarray} \nonumber
 I_E &=& -\left(1-\frac{\beta}{\beta_H}\right)\pi r_h^2 + \frac{M \beta}{2}-\frac{\beta r_h^3}{2 L^2}-\frac{\beta c_1 m^2 r_h^2 }{8}+\frac{\beta Q^2}{8 r_h}\\ \nonumber
 &=& \frac{\beta M}{2}+\frac{\beta r_h}{4}\Big[ 1+\frac{r_h^2}{L^2}+\frac{Q^2}{4 r_h^2}+m^2\left(\frac{c_1}{2}r_h+c_2\right)\Big]-\pi r_h^2\\
 &=& \beta M - S.
 \end{eqnarray}
To arrive at the final expression we have used Eq. \ref{3basics}. The corresponding partition function is given as,
 \begin{equation}
Z_{gravity}(\beta) \ \simeq e^{-\left(\beta M-S\right)}.
\end{equation}
 Now, the generalized free energy is readily obtained as,
 \begin{equation}\label{free_e}
 F = -T \ln Z_{gravity} = \frac{I_E}{\beta} =  \frac{r_h}{2} \Big[ 1+ \frac{r_h^2}{L^2}  + \frac{Q^2}{r_h^{2}}
+ \frac{c_1 m^2}{2} r_h 
+  c_2 m^2 \Big] - \pi T r_h^2.
\end{equation}
This expression for free energy is the same as the one obtained using standard thermodynamic relations \cite{Cai:2014znn, Xu:2015rfa} and, also by the Hawking-Witten prescription \cite{Dehghani:2019thq}. Once equipped with the generalized free energy, one can learn the stability and phase transition properties of black holes using various techniques as explained in \ref{intro}. The generalized free energy can also be expressed in terms of an effective thermodynamic pressure ($P$) in the extended phase space via the following identification \cite{Kastor:2009wy},
\begin{equation*}
P= \frac{3}{8 \pi L^2}.
\end{equation*}
The energy and entropy of the fluctuating black hole can be obtained easily from Eq. \ref{free_e} using conventional thermodynamic relations. As mentioned before, the free energy landscape for the black holes in massive gravity can be constructed by treating the horizon radius $r_h$ as the order parameter. The values of $r_h$ satisfying the relation $\partial G/\partial r_h =0$ represent different black hole phases. The black hole phase associated with the maxima of the free energy is thermodynamically unstable, whereas, the solutions corresponding to the minima are thermodynamically stable. The free energy landscape and the Hawking temperature profile of fluctuating black holes in massive gravity are given in Fig. \ref{fig_1}. On the left, the free energy landscape is plotted at the phase transition temperature where three black hole branches coexist. We distinguish different black hole phases according to the order parameter, i.e., small ($r_s$), intermediate ($r_m$), and large ($r_l$). One can easily deduce from the figure that the intermediate black hole is thermodynamically unstable whereas both small and large black hole phases are stable. On the right, Hawking temperature is plotted as a function of the order parameter. Three black hole states are labeled at the phase transition temperature.\\
The basic idea of the deterministic relaxation process of fluctuating black holes can be obtained from Fig. \ref{fig_1}. Such processes describe the deterministic evolution of the fluctuating black hole phase due to its interaction with the thermal bath.  Consider a fluctuating black hole with radius $r_f$, where $r_s<r_f<r_m$. From the plot of the ensemble temperature, one can see that the Hawking temperature of this fluctuating black hole is higher than the ensemble temperature. The deterministic relaxation process indicates that the fluctuating black hole reduces its temperature by giving out energy to the thermal bath. Since the mass of the black hole is a monotonic function of its radius, the fluctuating black hole will evolve to settle down to the stable small black hole phase. Note that the force due to fluctuation is not considered in this process. A detailed discussion on the stability and dynamics of phase transition of black holes in dRGT massive gravity using the free energy landscape is presented in \cite{Safir:2023thg}.\\
 
The deterministic gradient force  represents the deterministic tendency of the system to move towards lower energy states. In the case of black holes, it corresponds to the system's inclination to settle into local stable configurations, which are states with minimal free energy. However, in order to understand the dynamics of state switching one should consider the force due to fluctuations (thermal noises). The rest of the paper discusses the stochastic dynamics of black hole phase transition.

\section{Stochastic Dynamics of Black Hole Phase Switching} \label{second}
It is well known that the underlying force that is responsible for the black hole phase transition between two stable states is the thermal noise from the thermal bath surrounded by the black hole. Thermal noise arises due to the random motion of particles in the system, driven by temperature. This noise introduces fluctuations in the system's energy landscape, causing occasional deviations from deterministic behavior.  Naturally, to understand the dynamics of the phase-switching process one should resort to the tools of non-equilibrium statistical physics, i.e., the stochastic Langevin equation or equivalently the probabilistic Fokker-Planck equation. A detailed study of the dynamics of black hole phase transition in massive gravity using the Fokker-Planck equation is presented in \cite{Safir:2023thg}. In this section, we briefly outline the phase-space integral method to study the kinetics of the black hole phase-switching process using the Langevin equation.
\subsection{Stochastic Langevin Equation}
Langevin equation is used for finding the effect of fluctuations in the non-equilibrium systems. The equation contains both damping force and random forces, which are related according to the fluctuation-dissipation theorem. Both friction and noise arise from the interaction between the black hole and the thermal bath. In this approach, the fluctuations are introduced by adding random terms and are called stochastic noise sources \cite{Robert:2001zw,Kampen:2007ngv}. The Langevin equation for the dynamics of black hole phase switching can be written as \cite{Li:2024hje},
\begin{equation}
\ddot{\phi} + \zeta \dot{\phi}+ G'(\phi)-\bar{\eta}(t)=0.
\end{equation}
Here, $\phi$ denotes the order parameter. We follow the standard notation that $\dot{\phi} = \partial \phi/ \partial t$ and $G' = \partial G/\partial \phi$. The interaction between the black hole and its environment is characterized by an effective frictional coefficient $\zeta$. The parameter $\bar{\eta}$ denotes the stochastic noise which is independent of the order parameter. The stochastic characteristics of the noise parameter are given by specifying its first and second moment, i.e.,
\begin{equation}
<\bar{\eta}(t)>=0; \quad < \bar{\eta}(t)\bar{\eta}(t')>= 2 \zeta T \delta(t-t').
\end{equation}
The delta function reflects the fact that there is no correlation between the interactions at distinct time intervals $dt$ and $dt'$. In the over-damped regime (large $ \zeta$), the Langevin equation reduces to,
\begin{equation}\label{over_damp}
\dot{\phi} = -\frac{1}{\zeta} G'(\phi)+\eta(t),
\end{equation}
where $\eta(t) = \bar{\eta}/\zeta$. The correlation now becomes,
\begin{equation}
<\eta(t) \eta(t')> = 2 D \delta(t-t').
\end{equation}
This relation describes the stochastic motion of a Brownian particle with diffusion coefficient $D=T/\zeta$. As mentioned before, the path integral formalism can be used to describe non-equilibrium statistical systems to obtain useful insights. In the following section, we characterize the dynamics of black hole phase transition using MSRJD path integral formalism.
\subsection{MSRJD Path Integral Formalism}
 We briefly outline the derivation of MSRJD functional from the over-damped Langevin equation given in Eq. \ref{over_damp}. To describe a non-equilibrium system using path integral formalism, one usually employs the method of discretization, especially when the noise plays a significant role in the dynamics. We approximate a time continuous dynamics by a discrete-time process. However, the discretization can be done in various ways depending on the nature of the noise function. In Eq. \ref{over_damp}, the noise term $\eta(t)$ is realized as a series of delta peaks at random times and causes a jump in the value of $\phi$. Therefore, the value of $\phi$ at the arrival time of the delta function is undetermined. Naturally, one has multiple choices for opting for the value of $\phi$ during each step of the evolution. According to the Ito discretization scheme, the stochastic integral is defined such that the integrand is evaluated at the beginning of each time interval \cite{Kampen:2007ngv}. In this scheme, the order parameter ($\phi$) and the stochastic noise ($\eta$) are discretized as follows,
 \begin{equation}
 \phi(t) \rightarrow \phi_i; \quad \eta(t) \rightarrow \eta_i; \qquad i \in Z.
 \end{equation}
The over-damped Langevin equation becomes,
\begin{equation}\label{new_lang}
\psi_i:= \phi_i -\phi_{i-1}+\Delta t\left( \frac{1}{\zeta} G'(\phi_i)-\eta_i\right)=0,
\end{equation}
where $\Delta t$ is a discretized time interval. To derive the MSRJD functional, we consider the following identity,
\begin{equation}
\int \mathcal{D} \phi \ \delta  \left(\phi-\phi[\eta]\right) = \int \mathcal{D} \phi  \
 \Big|\frac{\delta \psi}{\delta \phi}\Big| \delta (\psi)=1.
\end{equation}
Here, $\phi[\eta]$ is a solution to the Langevin equation corresponding to the noise $\eta$. The functional integral measure $\mathcal{D}\phi = \Pi_{i} \ d\phi_i$ and $\delta(\phi-\phi[\eta]) = \Pi_{i}(\phi_1-\phi_i[\eta])$. Further, $ \Big|\frac{\delta \psi}{\delta \phi}\Big|$ is the determinant of the Jacobian matrix $\Big\{\frac{\partial \psi_i}{\partial \phi_j}\Big\}$. In the Ito discretization scheme, the determinant of the Jacobian matrix is one. Therefore, the above identity becomes,
\begin{equation}
\int \mathcal{D}\phi \ \prod_{i} \ \delta \left(\phi_i -\phi_{i-1}+\Delta t\left( \frac{1}{\zeta} G'(\phi_i)-\eta_i\right)\right)=1,
\end{equation}
where we have substituted for $\psi$ using Eq. \ref{new_lang}. Now, representing the delta function in terms of Fourier integral and taking the continuum limit, we get,
\begin{equation}
\int \mathcal{D}\phi  \ \mathcal{D}\bar{\phi} \ e^{-i\int dt \bar{\phi} \left(\dot{\phi} + \frac{1}{\zeta} G'(\phi) - \eta(t) \right)}   =1.
\end{equation}
The generating function for the Langevin dynamics in the over-damped regime is given by,
\begin{equation}
\mathcal{W} = \mathcal{N} \int \mathcal{D}\phi  \ \mathcal{D}\bar{\phi} \ \mathcal{D}\eta(t) \ e^{-\frac{1}{4D}\int dt \eta^2(t)} \ e^{-i\int dt \bar{\phi} \left(\dot{\phi} + \frac{1}{\zeta} G'(\phi) - \eta(t) \right)},
\end{equation}
where $\mathcal{N}$ denotes the normalization constant. Integrating over the noise gives,
\begin{equation}
\mathcal{W} = \mathcal{N} \int \mathcal{D}\phi  \ \mathcal{D}\bar{\phi} \ e^{-\int dt \big[i \bar{\phi}\left(\dot{\phi} + \frac{1}{\zeta} G'(\phi)\right) + D \bar{\phi}^2\big]}.
\end{equation}
The transition probability with initial state $\phi(t=t_0) =\phi_0$ and final state $\phi(t) = \phi$ is given by,
\begin{equation}
p(\phi,t;\phi_0,t_0) =  \int_{\phi_0}^{\phi} \mathcal{D}\phi  \ \mathcal{D}\bar{\phi} \ e^{-\int_{t_0}^t dt' \big[i \bar{\phi(t')}\left(\dot{\phi(t')} + \frac{1}{\zeta} G'(\phi(t'))\right) + D \bar{\phi}^2(t)\big]}.
\end{equation}
In the Hamilton formulation of the stochastic path integral, the probability can be expressed as,
\begin{equation}\label{final_action}
p(\phi,t;\phi_0,t_0) =  \int_{\phi_0}^{\phi} \mathcal{D}\phi  \ \mathcal{D}\bar{\phi} \ e^{-\int_{t_0}^{t} dt' \left(\Pi \dot{\phi}-\mathcal{H}\right)},
\end{equation}
where the conjugate momentum $\Pi = i\bar{\phi}$, and the effective Hamiltonian for the dynamics of state switching,
\begin{equation}\label{Hamilton}
\mathcal{H} = D \Pi^2 - \frac{1}{\zeta} G'(\phi)\Pi.
\end{equation}
Now, equipped with the MSRJD functional, we will study the kinetics of the black hole phase switching process in dRGT massive gravity. Note that the corresponding Fokker-Planck equation can be obtained easily from the above functional \cite{Li:2024hje}. 
\section{Hamiltonian Flow Lines and Kinetic Rate of Black Hole Phase Switching in Massive Gravity}\label{third}
Hamilton flow lines represent the trajectories that a system follows in phase space as it evolves according to Hamilton's equations. The Hamilton's equations can be readily obtained from Eq. \ref{Hamilton}.
\begin{eqnarray}\nonumber
\dot{\phi} \ &=& \ \frac{\partial \mathcal{H}}{\partial \Pi} \ = \ 2 D \Pi - \frac{G'(\phi)}{\zeta}\\
\dot{\Pi} \ &=& \ -\frac{\partial \mathcal{H}}{\partial \phi} \ = \ \frac{1}{\zeta} G''(\phi) \Pi
\end{eqnarray}
Now, we look for the fixed points in the phase space which reflect the stability of the system. Fixed points represent the state of dynamical systems where it does not evolve, i.e.,
\begin{equation*}
\dot{\phi} = \dot{\Pi}=0.
\end{equation*}
Thus, we have two sets of equations that correspond to the fixed points.
\begin{equation}
G'(\phi) = \Pi = 0; \quad \text{or} \ \quad G''(\phi)=0, \ \Pi = \frac{G'(\phi)}{2 T}.
\end{equation}
Here, we have expressed the diffusion constant in terms of temperature. The first set corresponds to zero Hamiltonian and represents stationary points on the free energy landscape. Also, these equations correspond to the on-shell black hole solutions with an extremum value of free energy. The second set of equations implies $G'(\phi) \neq 0$ and the corresponding Hamiltonian is negative. Therefore, we will not pursue it further. \\
\begin{figure}[hbt!]
    \centering
    \includegraphics[width=0.45\textwidth]{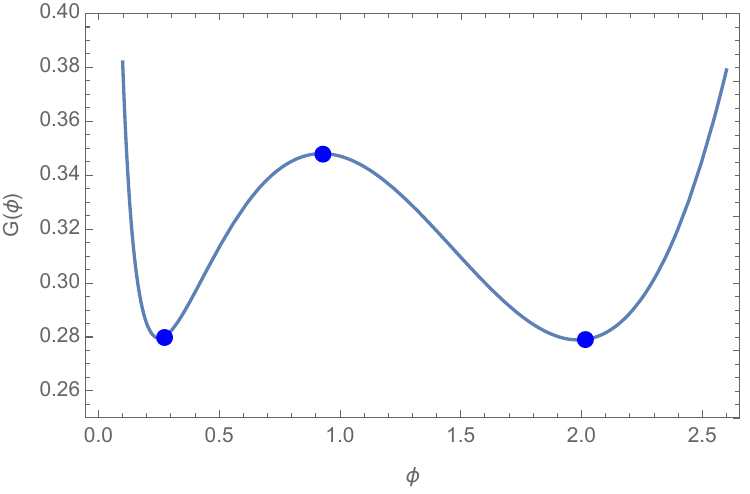}\;\;\;
    \includegraphics[width=0.45\textwidth]{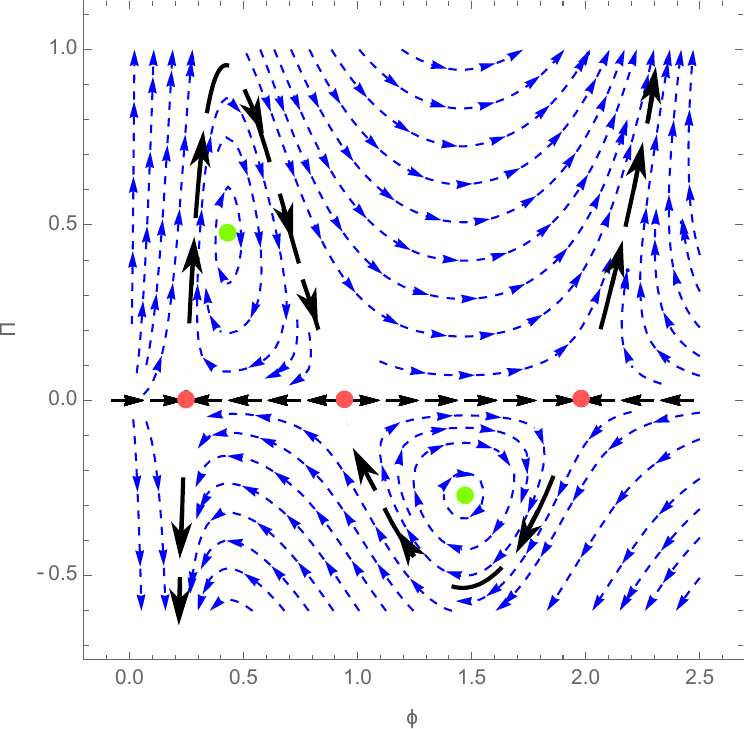}
      \caption{\textit{On the left:} The free energy landscape shows three black hole states at the phase transition temperature. The phases with small and large radii are stable whereas the intermediate one is unstable.   \textit{On the right:} The Hamiltonian flow lines for Hamilton's equations of motion are drawn. There are five fixed points, the saddle points are denoted by red points and the centers are in green. Each flow line on the plot is specified by a value of Hamiltonian. The black lines represent the flow lines corresponding to the vanishing Hamiltonian. The intersecting points between two zero flow lines correspond to three branches of black hole solutions. Numerical values used: $Q=0.5, P =0.03, m=1,c_1=0.02$, and $c_2=0.5$}
  \label{Flow_1}
\end{figure}

The free energy landscape and the Hamiltonian flow lines for black holes in dRGT massive gravity are given in Fig. \ref{Flow_1}. The two wells in the free energy landscape have equal depth because the ensemble temperature is taken to be the phase transition temperature. The Hamiltonian flow lines reflect the stability of the fixed points. The black arrows represent zero Hamiltonian lines and are described by,
\begin{equation}
\Pi=0 \quad \ \quad \text{or} \ \quad \ \quad \Pi = \frac{G'(\phi)}{T}.
\end{equation}
Along the $\Pi=0$ line, the Hamilton's equation reads,
\begin{equation}
\dot{\phi} = - \frac{1}{\zeta} G'(\pi).
\end{equation}
One can see from the figure that the solutions corresponding to the small and large order parameter states are stable whereas the state with the intermediate value of the order parameter is unstable along the $\Pi=0$ line. We have already observed this behavior from the free energy landscape, where, both small and large black holes have a minimum value of free energy. The intermediate black hole is at the local maxima of free energy and it would evolve away under perturbations. Now, the Hamilton equation corresponding to the $\Pi\neq0$ is given by,
\begin{equation}
\dot{\phi} = \frac{1}{\zeta} G'(\pi).
\end{equation}
Here, the nature of dynamic stability is reversed. The small and large black hole states are unstable whereas the intermediate one is stable.  Note that the Hamiltonian flow lines tell the dynamic nature of the system whereas the free energy landscape reflects the thermodynamic nature.\\

The intersections of the two zero energy flow lines at fixed points represent three branches of black hole solutions. We call these branches as small ($r_s$), intermediate $(r_m)$, and large $(r_l)$ states. The tunneling configuration for the black hole state switching can be easily understood from the plot. The state with the small radius ($r_s$) follows the Hamiltonian flow line represented by $\Pi \neq 0$ to reach the intermediate state ($r_m$). The intermediate black hole then follows the Hamiltonian flow line given by $\Pi =0$ to reach the large black hole state ($r_l$). Further, the reverse tunneling process can also be explained. The large black hole state would follow the Hamiltonian flow line characterized by $\Pi \neq 0$ to reach the intermediate state. The intermediate state then follows the $\Pi =0$ line to arrive at the small black hole state.\\

\begin{figure}[hbt!]
    \centering
    \includegraphics[width=0.45\textwidth]{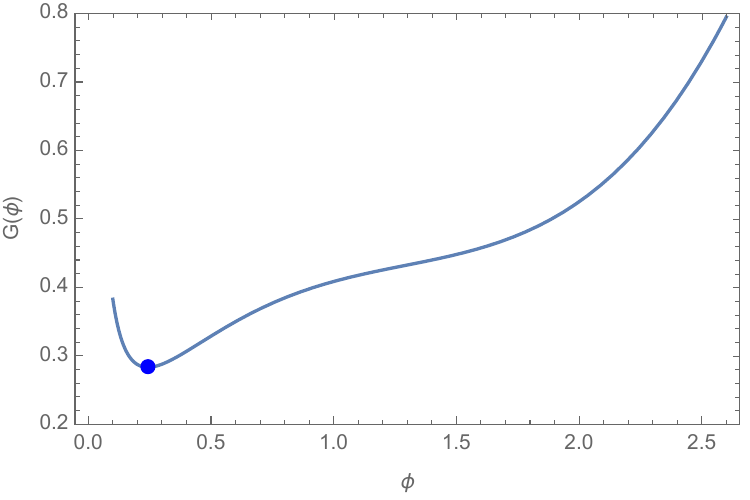}\;\;\;
    \includegraphics[width=0.45\textwidth]{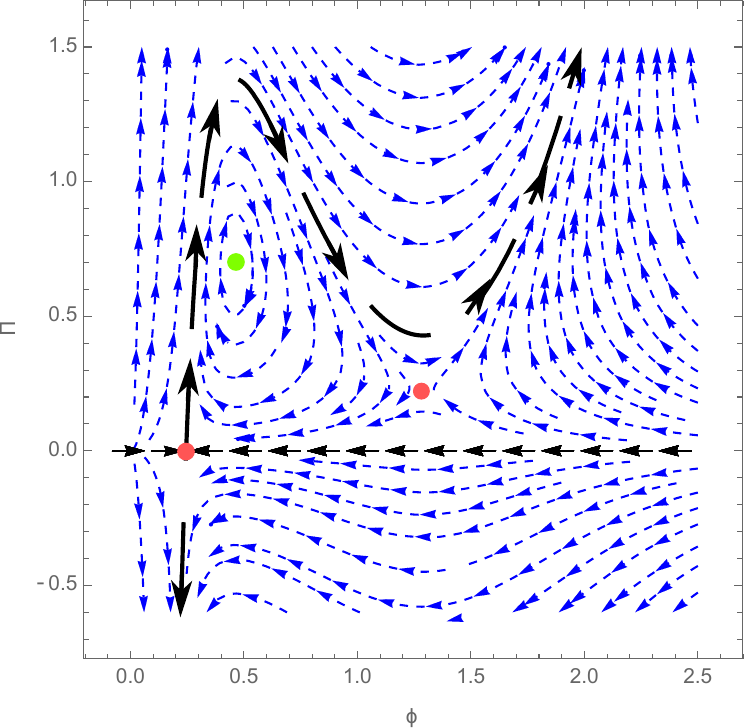}
      \caption{\textit{On the left:}  The single well configuration of the free energy landscape. The point at the minimum value of the free energy represents a stationary black hole.\textit{On the right:} The Hamiltonian flow lines with three fixed points. Saddle points are shown in red and the center point in green. The temperature is chosen such that the free energy landscape contains only one well. Numerical values used: $Q=0.5, P =0.03, m=1,c_1=0.02$, and $c_2=0.5$}
  \label{Flow_2}
\end{figure}

The single-well structure of the free energy landscape along with the Hamiltonian flow lines is plotted in Fig. \ref{Flow_2}. There exists a single black hole at the minimum value of free energy and is thermodynamically stable. There are three fixed points on the phase space including one center. However, there is only one saddle point that lies on the $\mathcal{H}=0$ flow line representing a small stationary black hole state. Therefore, tunneling configuration is absent in this case.\\

\begin{figure}[hbt!]
    \centering
    \includegraphics[width=0.45\textwidth]{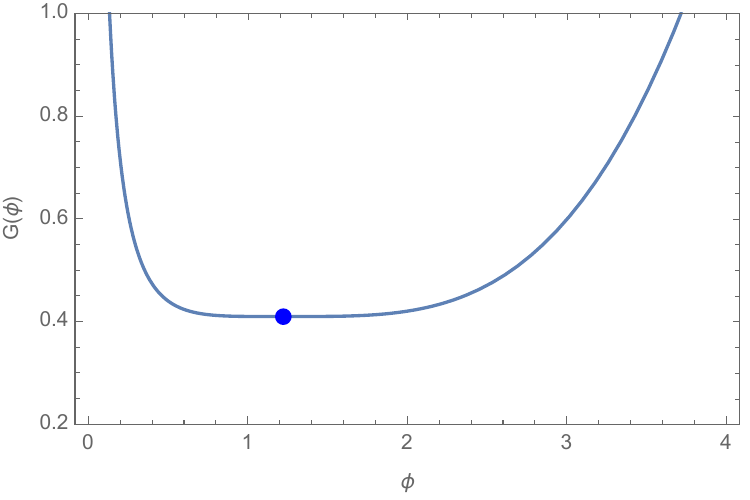}\;\;\;
    \includegraphics[width=0.45\textwidth]{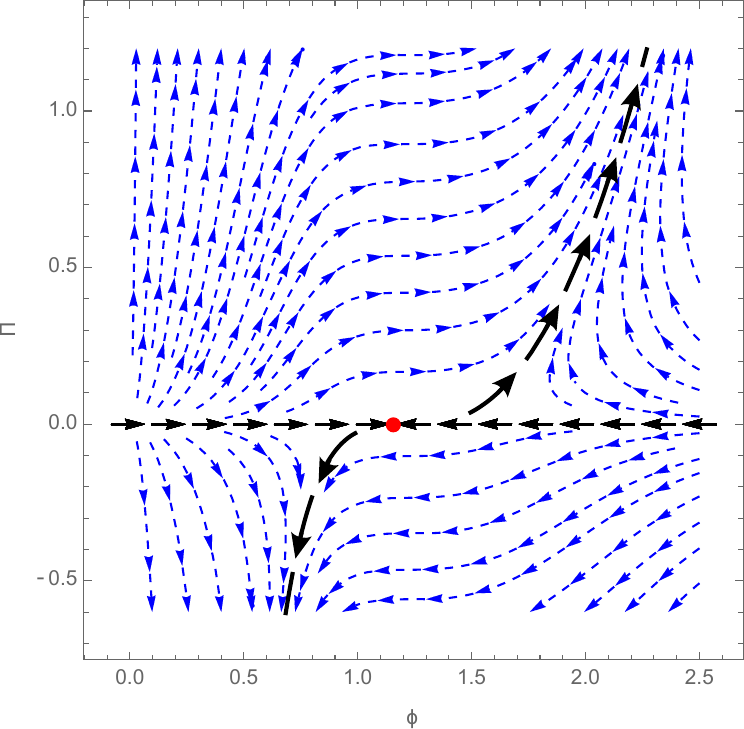}
      \caption{ \textit{On the left:} The free energy landscape is a single well containing a flat potential region. The blue point corresponds to a stable stationary black hole \textit{On the right:} The Hamiltonian flow lines contain only one fixed point. The fixed point is stable along $\mathcal{H}=0$ line but unstable along $\mathcal{H}\neq 0$ lines. Numerical values used: $Q=0.5, P_c =0.013, T_c=0.088$ and $m=c_1=c_2=0.2$}
  \label{Critt}
\end{figure}

\begin{figure}[hbt!]
    \centering
    \includegraphics[width=0.45\textwidth]{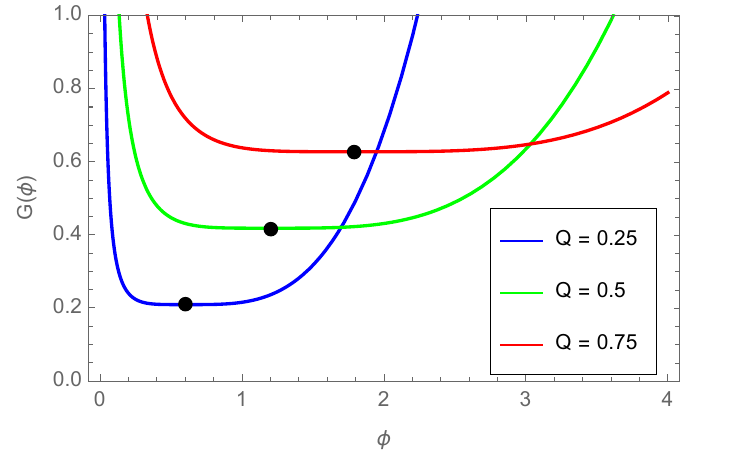}\;\;\;
    \includegraphics[width=0.45\textwidth]{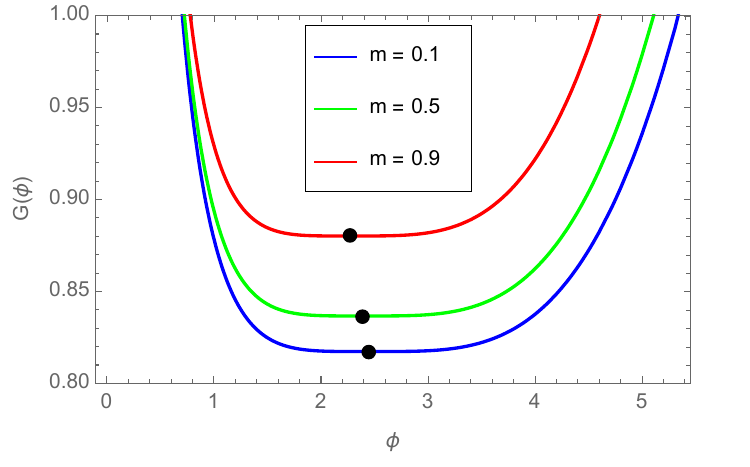}
      \caption{ \textit{On the left:} The single-well shape structure of the free energy landscape for different values of $Q$. The larger the black hole charge, the wider the well gets. \textit{On the right:} The free energy landscape corresponds to different graviton masses. The higher the graviton mass, the shape of the well gets sharper. The black point represents the critical value of the order parameter.}
  \label{effect}
\end{figure}

Now, we will study the characteristics of critical points in the phase space. Consider the free energy expression for  black hole states in dRGT massive gravity theory, now written in terms of $\phi$,
\begin{equation}
G(\phi)=\  \frac{\phi}{2} \Big[ 1+ \frac{\phi^2}{L^2}  + \frac{Q^2}{\phi^{2}}
+ \frac{c_1 m^2}{2} \phi 
+  c_2 m^2 \Big] - \pi T \phi^2.
\end{equation}
The critical point of evolution is characterized by the following conditions:
\begin{equation}
\frac{\partial G}{\partial \phi}\Big|_{\phi_c}=\ \frac{\partial^2 G}{\partial \phi^2}\Big|_{\phi_c}=\ \frac{\partial^3 G}{\partial \phi^3}\Big|_{\phi_c}=0.
\end{equation}
The corresponding values of order parameter$(\phi_c)$, temperature $(T_c)$, and pressure $(P_c)$ are obtained by solving the above set of equations and are given by,
\begin{eqnarray}\nonumber
\phi_c &=& \sqrt{\frac{6}{1+c_2m^2}} Q,\\
T_c &=& \frac{\left(1+c_2 m^2\right)^{\frac{3}{2}}}{3 \pi \sqrt{6} Q}\Big[1+\frac{3 \sqrt{6}Q c_1 m^2}{4\left(1+c_2 m^2\right)^{\frac{3}{2}}}\Big],\\ \nonumber
P_c &=& \frac{\left(1+c_2 m^2\right)^2}{96 \pi Q^2}.
\end{eqnarray}
These expressions match with the results presented in \cite{Hendi:2017fxp}. Note that the above expressions reduce to the critical values in Einstein's gravity theory when $m=0$ \cite{Li:2024hje}. We have plotted the free energy landscape and Hamiltonian flow lines at the critical point in Fig. \ref{Critt}. The free energy landscape takes the shape of a single well similar to the case described in Fig. \ref{Flow_2} except for the fact that the potential is flat in this case. The point on the curve corresponds to a stable stationary black hole solution. The Hamiltonian flow lines contain only one fixed point on the phase space and the tunnelling configuration is absent.
We observe that the structure of the kinetic path in configuration space and the kinetic rate of phase switching of black hole phase transition in dRGT massive gravity theory are qualitatively similar to the case of Einstein's gravity\cite{Li:2024hje}. The effect of black hole charge and the graviton mass on the free energy profile at the critical point is depicted in Fig. \ref{effect}. As the black hole charge increases the shape of the well gets wider. Also, the critical value of the order parameter increases as $Q$ increases. However, the graviton mass has the opposite effect on the free energy landscape. As the graviton mass increases, the shape of the well gets narrow and the critical value of the order parameter decreases. Finally, one can obtain the expression for the kinetic rate of phase switching  from the expression of action given in Eq. \ref{final_action} as,
\begin{equation}
k \sim e^{-\frac{\Delta G}{T}}.
\end{equation}
 Here, $\Delta G$ is the potential barrier between two black hole states and $T$ is the ensemble temperature. A detailed description of the probabilistic evolution of black hole phase transition in dRGT massive gravity is presented in \cite{Safir:2023thg}.
\section{discussion}\label{dis}
This paper has detailed the analysis of the black hole phase transition in dRGT massive gravity theory where the free energy is obtained from the gravitational partition function. The black hole is considered to be constantly interacting with the surrounding thermal bath exchanging energy. The event horizon radius is taken as the order parameter to describe the phase transition. We have derived an expression for Gibbs free energy from the gravitational partition function according to the Gibbons-Hawking method. The phase transition is caused by the thermal fluctuations due to which the black hole switches between small, intermediate, and large radii. The stability of different phases is explained by constructing the free energy landscape as a function of order parameters. We observe that the structure of the free energy landscape and the nature of stability of black hole phases in dRGT massive gravity theory are qualitatively similar to the charged black holes in Einstein's gravity theory\cite{Li:2024hje}.\\

The dynamics of phase transition is studied using MSRJD path integral formalism which is constructed from stochastic Langevin equations. Hamiltonian flow lines are constructed by solving Hamiltonian equations of motion and the dominant kinetic path corresponding to the phase transition on the phase space is analyzed. The dynamical stability of black hole states under phase transition is determined from the fixed points on the Hamiltonian flow lines. The Hamiltonian flow lines corresponding to the single-well free energy landscape are also considered. The critical behavior of black hole phase transition in dRGT massive gravity is studied and the critical values of the order parameter, temperature, and pressure are obtained. Finally, the effect of the black hole charge and the graviton mass on the critical behavior is analyzed. We observe that the width of the single-well shape structure of the free energy landscape is directly proportional to the black hole charge. Also, the critical value of the order parameter increases as the charge increases. On the other hand, the width of the well becomes narrower as the graviton mass increases.\\
It would be interesting to understand the construction of the generalized free energy from the gravitational partition function for higher dimensional dRGT massive gravity theory. We believe the thermodynamic structure and the phase transition properties in higher dimensional analysis will be very insightful because of the presence of more coupling constants. Also, another important extension of the present calculation will be to include the effect of Hawking radiation within the dynamics of phase transition. We leave these for future work.
\section{Acknowledgements}
This work is partially completed under Indian Academies' Summer Research Fellowship Program 2024. 
\appendix

\end{document}